\documentclass[twoside]{article}
\usepackage{balance}
\usepackage[sc]{mathpazo}
\usepackage[T1]{fontenc}
\usepackage{microtype}
\usepackage[hmarginratio=1:1,top=32mm,columnsep=20pt]{geometry}
\usepackage[font=it]{caption}
\usepackage{paralist}
\usepackage{multicol}
\usepackage{lettrine}
\usepackage{abstract}

\usepackage{titlesec}
\renewcommand\thesection{\Roman{section}}
\titleformat{\section}[block]{\large\scshape\centering}{\thesection.}{1em}{}
\usepackage{fancyhdr}
\usepackage{hyperref}

\title{\vspace{-15mm}%
	\fontsize{24pt}{10pt}\selectfont
	\textbf{Computer simulation control of single crystal growth process by melt pulling method}
	}	
\author{%
	\large
	\text{Jae Sik Jang, Un Chol Kye , Chol Jun Kang} \\[2mm]
	\normalsize	Department of Physics, \textbf{Kim Il Sung} university, Pyongyang, DPR Korea\\
		\vspace{-5mm}
	}
\date{}
\usepackage{makeidx}
\usepackage{latexsym}
\usepackage{amsfonts}
\usepackage{amssymb}
\usepackage{amsmath}
\usepackage{amsthm}
\usepackage{bbm}
\usepackage{mathrsfs}
\usepackage{color}
\usepackage{eucal}
\usepackage{caption}
\usepackage{cite}
\usepackage{mathdots}
\usepackage{graphicx,psfrag,subfigure}
\usepackage{amsbsy}
\usepackage{bbold}
\usepackage{dsfont}
\usepackage{textcomp}
\usepackage{comment}
\usepackage{indentfirst}
\usepackage{eso-pic}
\begin{document}

\maketitle
\thispagestyle{fancy}

\begin{abstract}
In this paper, on the basis of the set of simplified model state equations to represent the dynamic features of melt pulling method growth process, we constructed a simulation control system of Matlab Simulink and analyzed control features of state variables for the total growth process including shoulder growth process and the constant diameter growth process of single crystals such as Si and LiNbO3.
\end{abstract}

\begin{multicols}{2}
\lettrine[nindent=0em,lines=3]{A}utomatic control system of melt pulling method(Chochralski, CZ method) consists of systems for controlling the diameter of crystal. This is because the uniformity of crystal diameter is an important parameter determining the stability of crystal growth process and crystallographic or physical- chemical stability.
So far, various kinds of techniques such as weight measurement method of crystal or melt, measurement method of meniscus shape, measurement method of melt height ,etc., for automatically controlling the diameter of crystal, have been exploited, and computer simulations such as MC,MD,FEA ,etc., for optimizing the growth condition have been widely studied[\cite{1}-\cite{5}].
But those simulations have analyzed the crystal growth process only for static states because of the complexities of physico-chemical properties and the limited computer performance of CZ crystal growth system. Thus, they could not exactly take into account the dynamic process of crystal growth and it is difficult to do real-time control.
Recently, studies on computer simulation control such as PID control, model prediction control(MPC) for optimizing the control of crystal growth were developed, but they could not exactly represent realistic feature of CZ crystal growth system due to mismatching between model and control system[\cite{6}-\cite{13}].

\section{State equations of CZ crystal growth process}
\begin{figure*}
\centering
\includegraphics[width=0.7\textwidth]{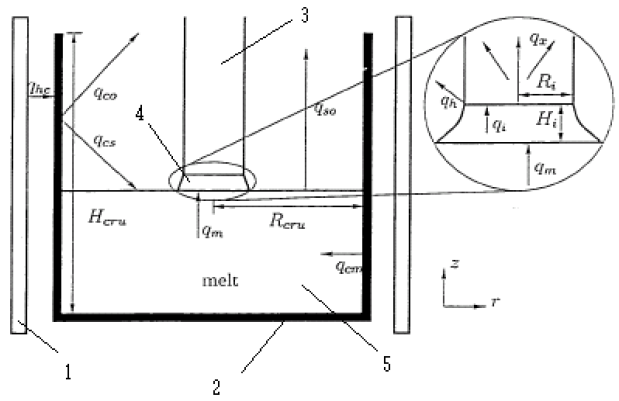}
\hspace{20mm}
\caption{Simplified model of crystal growth system (1-heater,2-crucible,3-growing crystal,4-meniscus,5-melt)}
\label{Fig1}
\end{figure*}
CZ crystal growth system is modeled as shown in Fig.(\ref{Fig1}). State equations to represent the relation between state variables($T_h$,$T_c$,$T_m$,$R_i$,$H_i$,$R_e$,$H_m$,$\phi$) and input variables($V_p$, $P_{in}$) are classified into temperature and geometric state equations[\cite{14}].
\subsection{Set of temperature state equations}
The temperature state variables of the model are temperature of   heater $T_h$, temperature of crucible $T_c$ and temperature of melt $T_m$ . Time derivative of temperature of heater is expressed as follows:
\begin{equation}
\frac{dT_h}{dt}=\frac{1}{C_h}(P_{in}-q_{hc})
\end{equation}
, where $C_h$ is heat capacity of heater, $P_{in}$ is input power of heater, $q_{hc}=A_c \sigma (T^4_h-T^4_c)$ is heat radiation transport rate from heater to crucible, $A_c$ is surface area of crucible and $\sigma$ is Sthefan Boltzmann constant.
Time derivative of temperature crucible is
\begin{equation}
\frac{dT_c}{dt}=\frac{1}{C_c}(q_{hc}-q_{co}-q_{cs}-q_{cm})
\end{equation}
where $C_c$ is heat capacity of crucible, $q_{co}$ is heat radiation transport rate toward environment, $q_{cs}$ is heat radiation transport rate toward melt and $q_{cm}$ is heat conduction transport rate toward melt. Time derivative of temperature of melt is
\begin{equation}
\frac{dT_m}{dt}=\frac{1}{C_m}(q_{cm}+q_{cs}-q_{so}--q_{m})-\frac{\dot{H}_m T_m}{H_m}
\end{equation}
where $C_m$ is heat capacity of melt, $\dot{H}_m$ is derivation of melt height with time, $q_m$ is heat transport  rate from melt to meniscus.
\subsection{Set of geometric state equations}
The geometric state variables of model are radius of  crystal $R_i$, meniscus height of melt $H_i$, height of melt $H_m$, meniscus contact angle $\phi$ and effective radius of crystal $R_e$. The time derivative of height of melt can be written from the condition of mass equilibrium at crystal/melt interface as
\begin{equation}
\frac{dH_m}{dt}=-\frac{\rho R^2_i (V_p-\dot{H}_i)}{\rho_l R^2_e-\rho_s R^2_i}
\end{equation}
where $V_p$ is crystal pulling rate, $\dot{H}_i$ is differentiation meniscus height with time, $\rho_s$ and $\rho_l$ are densities of crystal and melt respectively.
From thermal equilibrium condition at the crystal/melt interface we have
\begin{equation}
\frac{dH_i}{dt}=\frac{\rho_l R^2_c V_p}{\rho_l R^2_c-\rho_s R^2_i}-\frac{q_x-q_i}{H_f \rho_s \pi R^2_i}
\end{equation}
where $H_f$  is crystallization latent heat, $q_x$ is heat transport rate from crystal/melt interface toward crystal and $q_i$ is heat transport rate from meniscus toward crystal/melt interface.
Taking account of the conditions for thermal and mass equilibrium and Laplace-Young equation, time variation of crystal radius and meniscus contact angle is 
\begin{equation}
\frac{dR_i}{dt}=\frac{q_x-q_i}{H_f \rho_s \pi R^2_i}\tan{(\phi-\phi_0)}
\end{equation}
\begin{equation}
\begin{array}{cc}
\frac{d\phi}{dt}=&-\left[ 4 \left( \frac{\rho_l R^2_{cru} v_p}{\rho_l R^2_{cru} -\rho_s R^2_l}-\frac{q_x-q_i}{H_f \rho_s \pi R^2_i} \right) S_i R^2_i \right.  \\
&\left. +\frac{\beta \cos{\phi} (q_x-q_i) \tan{(\phi-phi_0)}  (\beta \cos{\phi}-S_1)}{H_f \rho_s \pi R^2_i}\right]/D_{phi}
\end{array}
\end{equation}
where $\phi$ is meniscus contact angle of melt, $\phi_0$ is the material dependent meniscus contact angle that produces constant diameter crystal growth, $S_1=\sqrt{\beta \left( 16 R^2_i(1-sin\phi)+\beta cos^2 \phi  \right)}$ and $\beta$ is Laplace constant. 
The time derivative of effective radius of crystal is expressed as follows:
\begin{equation}
\frac{dR_e}{dt}=\frac{1}{\tau}(R_i-R_e)
\end{equation}
where $\tau=f_m R_i/V_p$ is average time and $f_m$ is about 0.25$\sim$0.5.
\section{Control features of model based on the state equations}
\subsection{Si single crystal growth process}
\begin{figure*}
\centering
\includegraphics[width=0.7\textwidth]{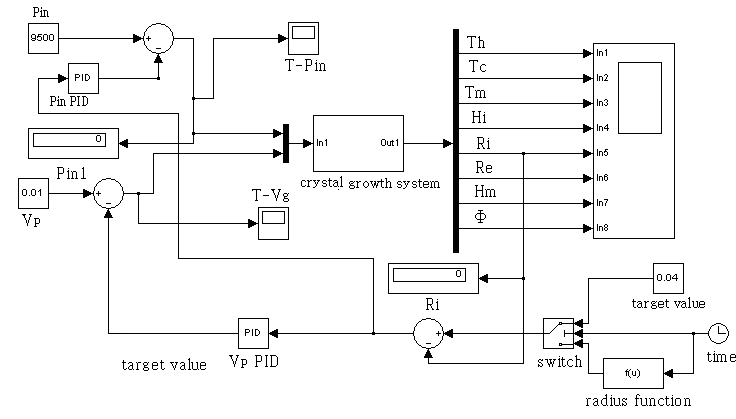}
\caption{Schematic diagram of simulation controller}
\label{Fig2}
\end{figure*}
Simulation control was made for the growth process of typical semiconductor crystal Si with 3-inch diameter. The structure of simulation controller is shown in Fig.(\ref{Fig2}).
The simulation control results of Matlab Simulink are shown in Fig.(\ref{Fig3}).
\subsection{LiNbO3 single crystal growth process}
\begin{figure*}
\centering
\includegraphics[width=0.8\textwidth]{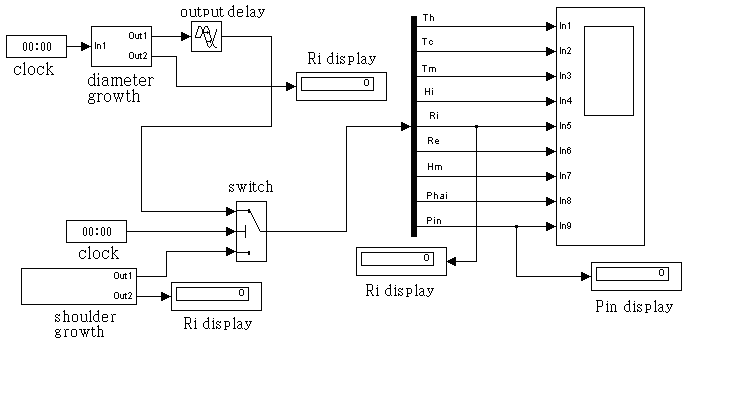}
\caption{Schematic diagram of simulation controller}
\label{Fig4}
\end{figure*}

\begin{figure*}
\begin{center}
    \begin{tabular}{cc}
     \resizebox{55mm}{!}
{\includegraphics{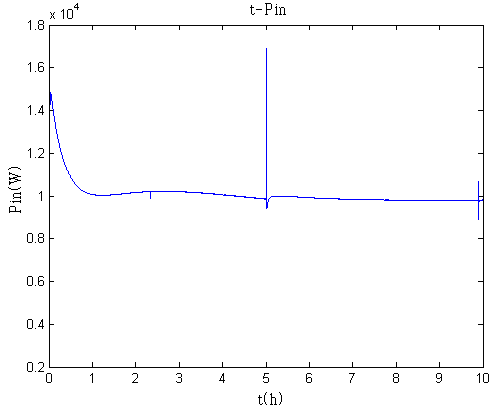}} &
     \resizebox{60mm}{!}
{\includegraphics{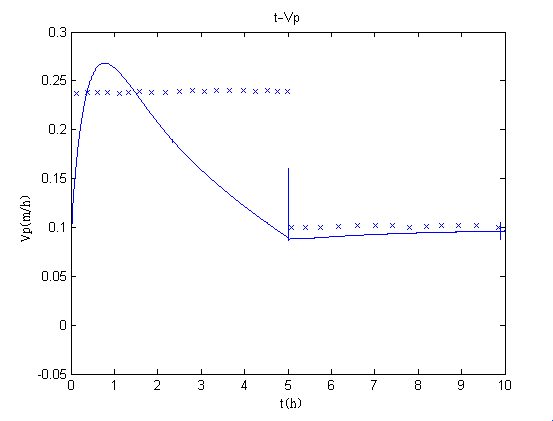}}\\
 \resizebox{55mm}{!}
{\includegraphics{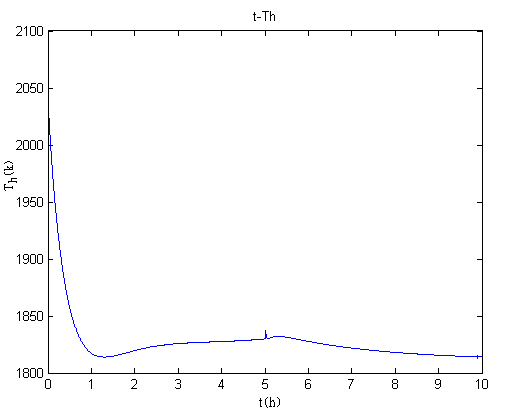}} &
     \resizebox{55mm}{!}
{\includegraphics{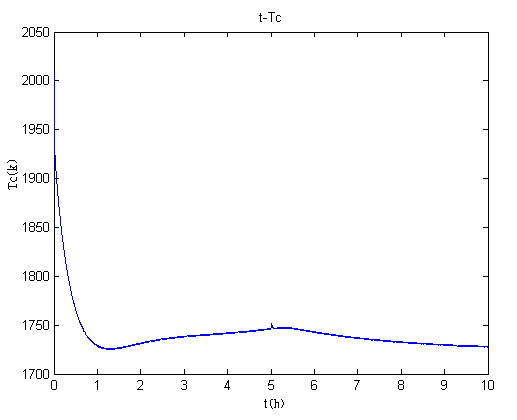}}\\
 \resizebox{55mm}{!}
{\includegraphics{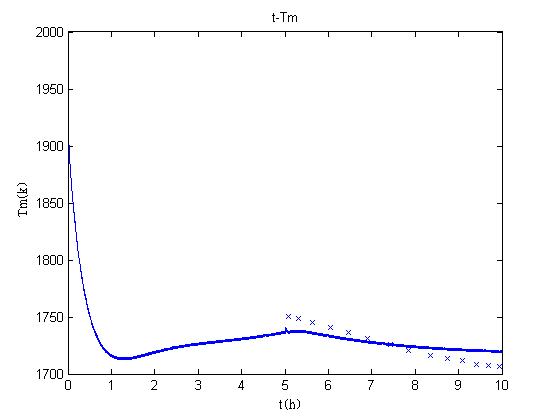}} &
     \resizebox{55mm}{!}
{\includegraphics{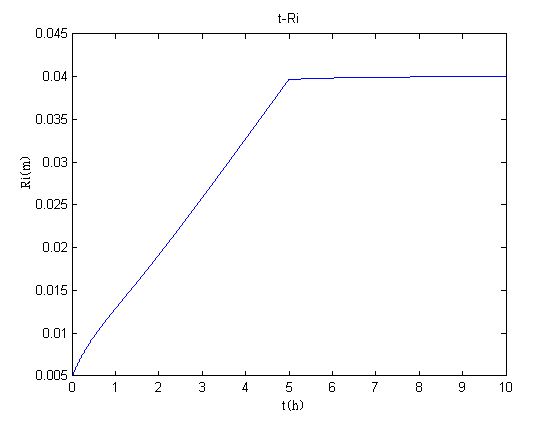}}\\
 \resizebox{55mm}{!}
{\includegraphics{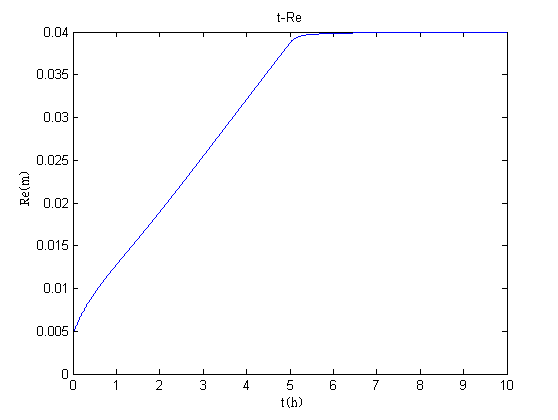}} &
     \resizebox{55mm}{!}
{\includegraphics{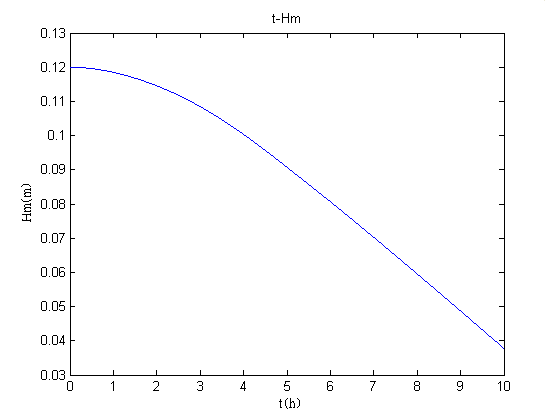}}\\
 \resizebox{55mm}{!}
{\includegraphics{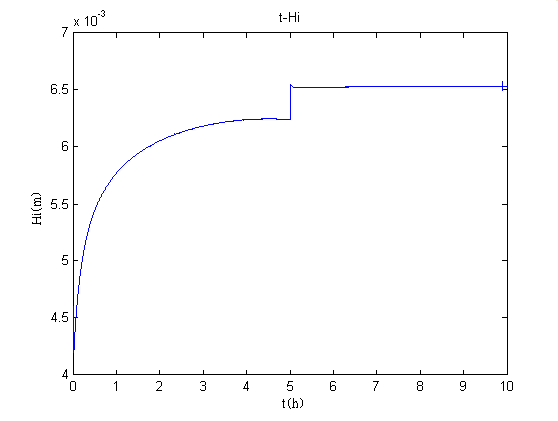}} &
     \resizebox{55mm}{!}
{\includegraphics{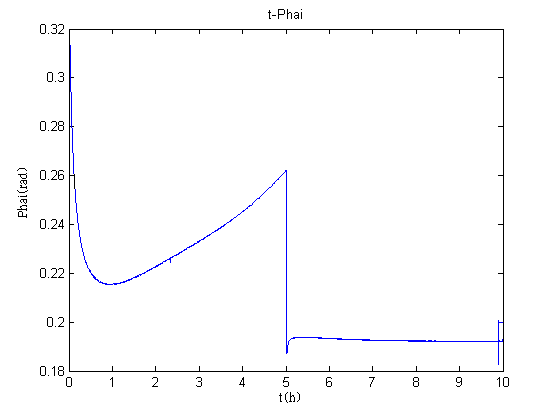}}\\
\end{tabular} 
 \caption{Plot of state variables versus growth time ($\times$ :  Experimental data)}
   \label{Fig3}
\end{center}
\end{figure*}

From the growth process features of oxide single crystal, controller can be made of two independent simulation parts : i.e. simulation controllers of shoulder growth process and constant diameter growth process (Fig.(\ref{Fig4})). 
The simulation control results of Matlab Simulink are shown in Fig.(\ref{Fig5}).
\section{Discussion}
\subsection{Si single crystal growth process}
Control feature of shoulder growth step of crystal has a number of differences in comparison with constant diameter growth step. Input power $P_{in}$ varies in the range of magnitude above $10^4W$, then slowly decreases in constant diameter growth step. Pulling rate $V_p$ is also $3\sim4㎜/min$ in the shoulder growth step and $1.5\sim2mm/min$ in constant diameter growth step.
Meniscus contact angle is $20$$\sim$$30^{\circ}$($\phi>\phi_0$) in the shoulder growth step and is maintained in constant diameter growth step as $11^{\circ}$ ($\phi=\phi_0$)[\cite{5}]. Meniscus height is low in shoulder growth step and maintained at 6.176mm in constant diameter growth step. $T_h$, $T_c$ and $T_m$ decrease within about 200K and then slowly increases in shoulder growth step, thereafter decrease smoothly in constant diameter growth step.
This is in good agreement with ref.[\cite{15}-\cite{17}] related to growth simulation. The size of Si single crystal obtained by simulation controller is $\phi 8cm \times l13cm$. The $q_i/q_x$  is 0.6 in the shoulder growth step and is 0.8 in the constant diameter growth step.
\subsection{LiNbO3 single crystal growth process}
\begin{figure*}
\begin{center}
    \begin{tabular}{cc}
     \resizebox{55mm}{!}
{\includegraphics{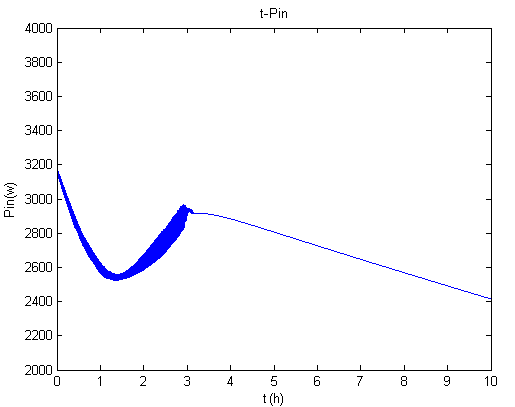}} &
     \resizebox{55mm}{!}
{\includegraphics{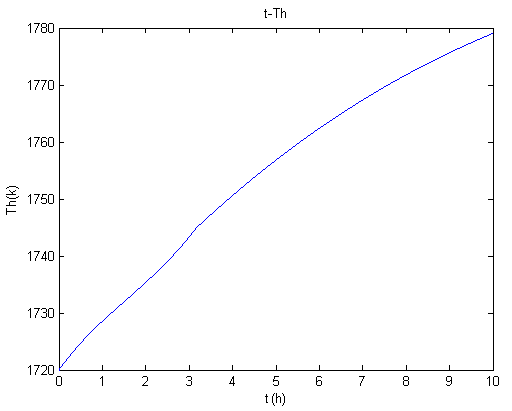}}\\
 \resizebox{55mm}{!}
{\includegraphics{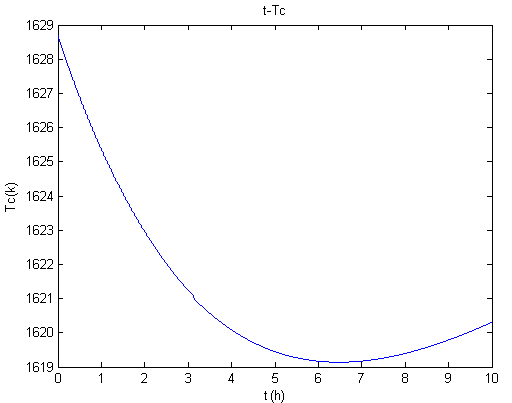}} &
     \resizebox{55mm}{!}
{\includegraphics{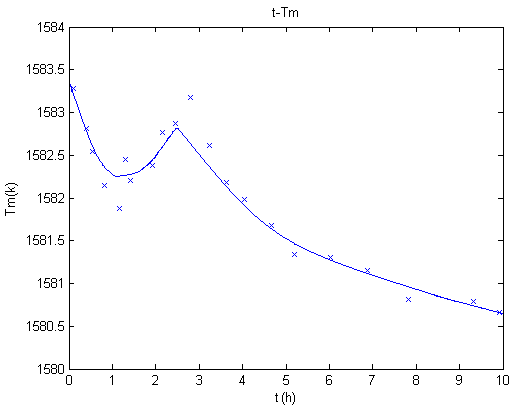}}\\
 \resizebox{55mm}{!}
{\includegraphics{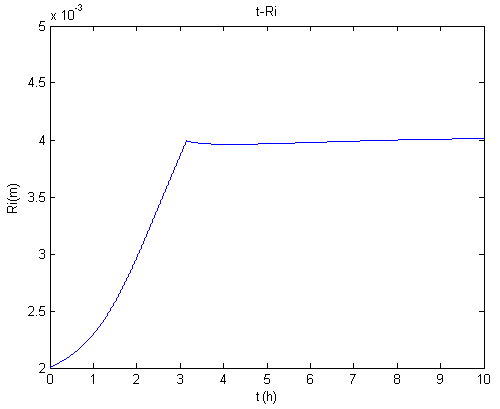}} &
     \resizebox{55mm}{!}
{\includegraphics{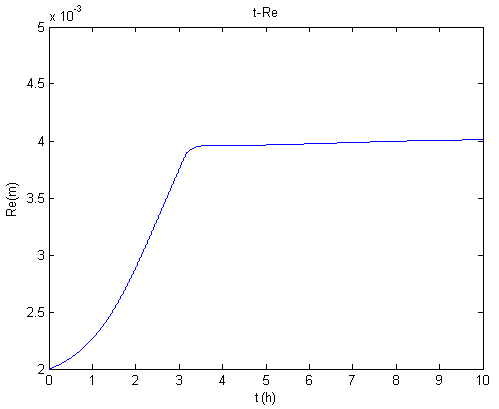}}\\
 \resizebox{55mm}{!}
{\includegraphics{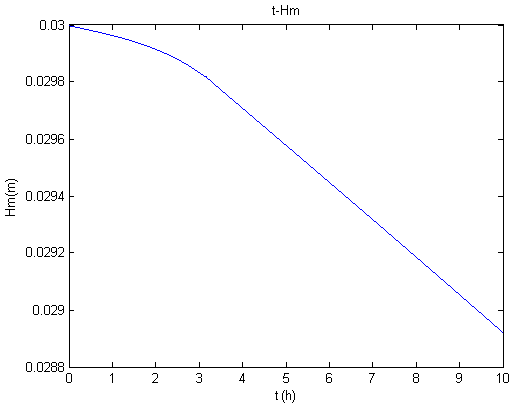}} &
     \resizebox{55mm}{!}
{\includegraphics{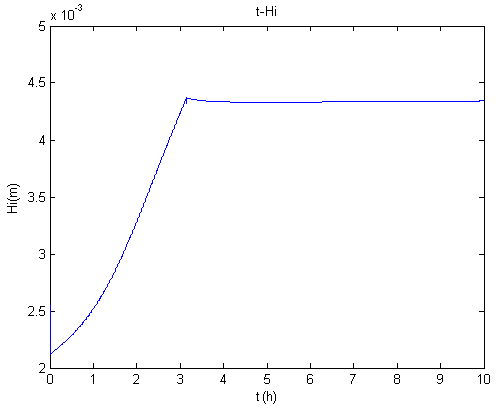}}\\
 \end{tabular} 
\centering
\includegraphics[width=0.4\textwidth]{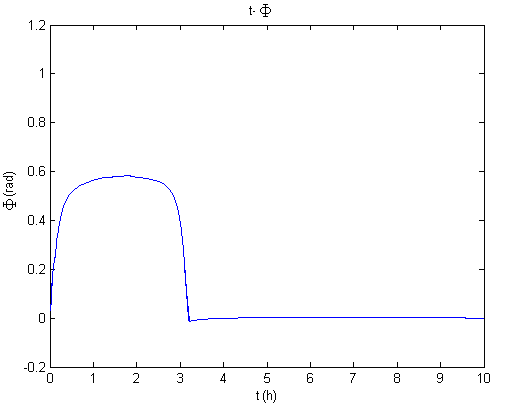}
 \caption{Plot of state variables versus growth time ($\times$ : Experimental data\cite{19})}
 \label{Fig5}
\end{center}
\end{figure*}

Fluid dynamical properties of melt for semiconductor crystals and oxide crystals are considerably different and hence pulling rate in semiconductor crystal growth is the order of a few mm/min, but in oxide crystal growth being the order of more than 1/10 of it.
Because of these properties of melt, in semiconductor crystal growth not only input power but pulling rate  is  controlled to obtain single crystal with desired diameter,but in oxide crystal growth pulling rate can not be controlled. Therefore, only input power can be controlled in oxide crystal growth.
Input power $P_{in}$ decreases from magnitude of 3200W to 2500W and then again increases up to 2800W during shoulder growth process, while being considerably fluctuated with the amplitude of 20W. It is slowly decreased from 2800W to 2400W with the rate of 55W/h in the constant diameter growth step. This is in good agreement with experimental data [\cite{18}]. Meniscus height $H_i$ increases from 2mm up to 4.3mm and then maintained at the value for steady state. Meniscus contact angle $\phi$ varies following the law to represent $\phi>\phi_0$ for shoulder growth step and $\phi=\phi_0$ for constant diameter growth step [\cite{5}].
Temperature of heater $T_h$ increases about 40K with the rate of 4K/h , temperature of melt $T_m$ initially decreases about 2K at 1583.2K and then increases up to 1583K, thereafter decreases down to 1581K smoothly(Fig.(\ref{Fig5})). This is in good agreed with experimental data [\cite{19}].
The size of LiNbO3 single crystal obtained by simulation controller is $\phi 8mm\times l 6cm$. The ratio of $q_i/q_x$   is 0.2 in the shoulder growth step, and is 0.6 in the constant diameter growth step.
\section{Conclusions}
On the basis of establishing the set of state equations of simplified model for CZ growth process, we have constructed the simulation controller that can control the total process including shoulder growth and constant diameter growth of crystal and by using it, we have controlled the growth process of single crystals such as Si and LiNbO3 to maintain the uniformity of crystal diameter.
Results of simulation control have shown that each state variable expresses the process of growth exactly. Thus, the simulation control system of CZ method based on simplified model suggests the methodology and theoretical basis by which selecting any state variable as a controlled object, one can design a program-control system for the growth process of single crystal with constant diameter with relatively low cost.

%
\end{multicols}

\end{document}